# SIMULATION OF A HIGH-ENERGY ELECTRON BEAM TRANSMISSION THROUGH TITANIUM AND KAPTON® THIN FILMS

Tetiana V. Malykhina[a,b,*], Stepan G. Karpus[b], Oleg O. Shopen[b], Valerii I. Prystupa[b]

[a]*Kharkiv V.N. Karazin National University*
*4, Svobody sq., 61022, Kharkiv, Ukraine*
[b]*National Science Center "Kharkiv Institute of Physics and Technology"*
*2, Academicheskaya str., 61108, Kharkiv, Ukraine*
*\*Corresponding Author: malykhina@karazin.ua*

The results of computer simulation of the high-energy electrons passage through thin layers of titanium (Ti) and polyimide Kapton® ($C_{22}H_{10}N_2O_5$) in the energy range from 3 MeV to 20 MeV are presented. Simulation is carried out using the Geant4 toolkit. The number of primary electrons is $6.24 \times 10^7$ for each series of calculations. The thickness of the titanium foil in the model experiment is 50 μm, the thickness of the Kapton® film is 110 μm. The energies of primary electrons are chosen as following: 3 MeV, 5 MeV, 10 MeV, 15 MeV, and 20 MeV. The purpose of the calculations is to reveal the possibility of using the Kapton® film in the output devices of linear electron accelerators. It was necessary to calculate the probable values of the energy absorbed in a Kapton® film and in a titanium foil for each value of primary electrons energy. Another important characteristic is the divergence radius of the electron beam at a predetermined distance from the film, or the electron scattering angle. As a result of calculations, the energy spectra of bremsstrahlung gamma-quanta, formed during the passage of electrons through the materials of the films, are obtained. The most probable values of the energy absorbed in the titanium foil and in the Kapton® film are calculated. The scattering radii of an electron beam for the Kapton® film and also for the titanium foil at a distance of 20 centimeters are estimated. These calculations are performed for electron energies of 3 MeV, 5 MeV, 10 MeV, 15 MeV, and 20 MeV. A comparative analysis of the obtained results of computational experiments is carried out. It is shown that the ratio of the total amount of bremsstrahlung gamma quanta in the case of use the Kapton® film is approximately 0.56 of the total amount of bremsstrahlung gamma quanta when using the titanium foil. The coefficients of the ratio of the electrons scattering radius most probable value after passing through Kapton® to the most probable value of the scattering radius after passing through titanium are from 0.62 at electrons energy of 3 MeV to 0.57 at electrons energy of 20 MeV. The analysis of the calculated data showed that the use of Kapton® ($C_{22}H_{10}N_2O_5$) as a material for the manufacture of output devices for high-energy electron beams is more preferable in comparison to titanium films, since the use of Kapton® instead of titanium makes it possible to significantly reduce the background of the generated bremsstrahlung gamma quanta and reduce the scattering radius of the electron beam.

**Keywords:** bremsstrahlung, Geant4-simulation, LINAC, a Kapton® film, ($C_{22}H_{10}N_2O_5$), interaction of radiation with matter, electron angular scattering.
**PACS:** 07.05.Tp, 02.70.Uu, 81.40wx

Linear electron accelerators are used to solve various applied problems related to the processes and effects occurring during the interaction of ionizing radiation with matter. These problems include irradiation of samples required for research in nuclear and medical physics, in particular, the improvement of methods for the production of radioisotopes. Applied studies of irradiated materials properties, as well as many other problems, are no less important tasks. All these tasks in most cases require an electron beam with certain characteristics. For example, there is a requirement to minimize the radius of the electron beam, as well as to reduce the bremsstrahlung background after passing through the output device of the electron accelerator. Great attention is paid to the study of the influence of the primary electron flux angular distribution on the irradiated object in work [1]. In particular, the study of the absorbed dose distribution was carried out depending on the penetration depth of the electron beam. The electron scattering angle at low energies (up to 10 MeV) was calculated for polyethylene layers. The layers thickness was more than 1 radiation length. The results of these studies suggest the expediency of studying the angular scattering of electrons after passing through thin polymer films, which could be used in the output devices of accelerators. We chose Kapton® ($C_{22}H_{10}N_2O_5$) as an object to study the possibility of optimizing the accelerator beams parameters in the energy range from 3 MeV to 20 MeV. We took into account the physical and chemical properties [2, 3] of Kapton®, in particular, the Kapton® density, the value of the critical energy, etc. Kapton® is a fairly stable material in the range from low temperatures to +400°C [4]. Despite the widespread use of Kapton® [3, 4] in medicine, aircraft construction, astronautics, vacuum technology, and other industries, the possibility of its use in accelerator technology has not been sufficiently studied.

## MATERIALS AND METHODS

We carried out a computer simulation of the passage of electron beams of various initial energies through a titanium foil. This simulation is carried out in order to study the possibility of optimizing the parameters of the output devices of linear electron accelerators. The output window of a typical LINAC usually is made of titanium foil. We consider the titanium foil thickness of 50 microns. This value of the titanium foil thickness is selected in accordance to the real foil thickness used in the LINAC-300 accelerator at National Scientific Center "Kharkiv Institute of Physics and Technology". The energies of primary electrons are equal to 3 MeV, 5 MeV, 10 MeV, 15 MeV, and 20 MeV in the simulation. The fluencies of bremsstrahlung gamma quanta and the electron beam scattering radii are calculated at a distance of 20 centimeters after the foil, since an irradiated target is supposed to be installed in this position. Similar calculations are performed for the Kapton® film with a thickness of 110 μm. This thickness of the Kapton® film was chosen for our research because of studies of the mechanical properties of various thicknesses of Kapton® films were carried out earlier in laboratory conditions. The obtained data were in good agreement with the data from the manufacturers' catalogs [5], as well as with the literature data [6].

We have developed a computer program in the C++ language that uses the Geant4 toolkit [7–9] for modeling the processes of interaction of radiation with matter. The primary electrons amount is $6.24\times10^7$, and the threshold $E_{cut}$ for particle tracking [7] is 0.1 μm. We have used the low energy `emlivermore` model of the `PhysicsList` unit. This model is based on the data of EEDL, EPDL libraries [9], and is applicable for electromagnetic processes modeling in the energy range from 100 eV to 20 GeV.

The results of Monte Carlo simulation of the absorbed energy spectra in the titanium foil and in the Kapton® film are shown in Figure 1. The calculation results are normalized to 1 primary electron. It can be seen that the most probable value of the absorbed energy in the case of using Kapton® is slightly less than the most probable value of the absorbed energy in the case of using titanium. The value of the absorbed energy is calculated for two values of the primary electrons energy. These values are 3 MeV and 15 MeV. The most probable values of the absorbed energy in Kapton® and titanium are obtained as a result of statistical data processing. The value of the electron energy absorbed in Kapton® is 17 keV for both values of the primary electron energy. The most probable energy absorbed in titanium is 20 keV for primary electrons with energy of 3 MeV, and 21 keV for primary electrons with an energy of 15 MeV. The statistical error is no more than 1%. The step of creating histograms when calculating the energy spectra is 1 keV.

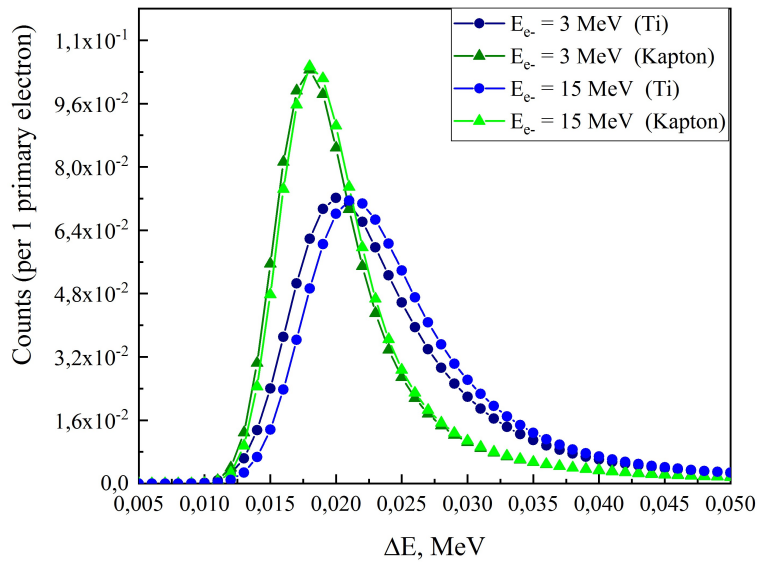

**Figure 1.** Energy spectra of absorbed energy in the titanium foil and in the Kapton® film for primary electron energies of 3 MeV and 15 MeV (triangles indicate values for Kapton®, circles indicate values for titanium)

The total and average values of the energy absorbed in titanium and in Kapton® during the passage of 3 MeV electrons beam, as well as 15 MeV electrons beam are presented in Table 1.

**Table 1.**
Total (for $N_e=6.24\times10^7$) and average values of energy absorbed in titanium and in Kapton®

| Titanium 50 microns | | |
|---|---|---|
| Electrons energy, MeV | Total absorbed energy, MeV | Average value of absorbed energy, MeV |
| 3 | Sum= $1.79\times10^6$ | Sum/$N_e$= 0.0286 |
| 15 | Sum= $1.82\times10^6$ | Sum/$N_e$= 0.0292 |
| Kapton® 110 microns | | |
| Electrons energy, MeV | Total absorbed energy, MeV | Average value of absorbed energy, MeV |
| 3 | Sum=$1.514\times10^6$ | Sum/$N_e$= 0.0243 |
| 15 | Sum=$1.514\times10^6$ | Sum/$N_e$= 0.0243 |

The scattering radii of electrons at a distance of 20 cm after passing through a 50 μm thick titanium foil are estimated by the Monte Carlo method. A similar series of simulations is carried out to estimate the scattering radii of electrons after passing through a Kapton® film with a thickness of 110 μm. The distance of 20 cm after the foil is chosen as the position at which we estimate the electron scattering radius, because the irradiated sample will be placed at this position. The calculations are carried out for several values of the primary electrons energy. These values are 3 MeV, 5 MeV, 10 MeV, 15 MeV, 20 MeV. The number of simulated events is $6.24\times10^7$. The graphs characterizing the value of the scattering radius of the electron beam after passing through the titanium foil, as well as after passing through the Kapton® film, for each energy of primary electrons are shown in Figure 2. The corresponding values of the scattering radii, as well as the comparison result, are presented in Table 2. The error in determining the scattering radii of electrons is 0.1 mm. This error is due to the size of the histograms calculating step during data processing. Graphs of changes in the values of the electrons scattering radii depending on the primary electrons energy for each of two materials are shown in Figure 3.

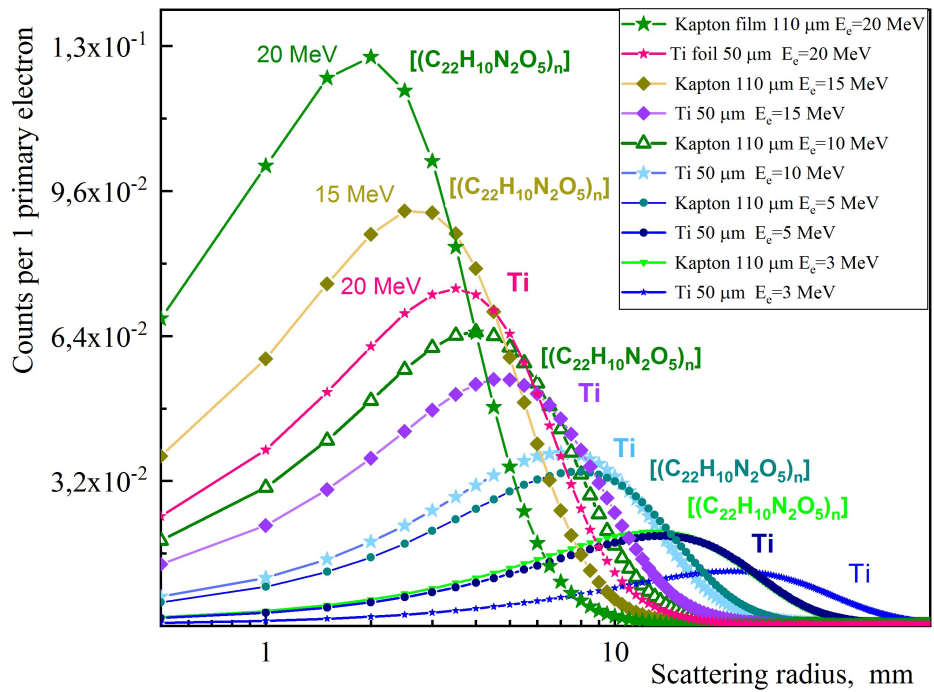

**Figure 2.** The scattering radius of electrons of various energies after passing through the titanium foil as well as after the Kapton® film at 20 cm distance after the film or foil

**Table 2.**
The most probable values of scattering radii of electrons at the distance of 20 cm after the titanium foil or the Kapton® film, as well as the ratio coefficient of the electron scattering radii

| Electrons energy, MeV | The most probable scattering radius (mm) of electrons after the Kapton® film, 110 μm | The most probable scattering radius (mm) of electrons after the Ti foil 50 μm | The coefficient of the scattering radii ratio |
|---|---|---|---|
| 3 | 13 | 21 | 0.62 |
| 5 | 8 | 14 | 0.57 |
| 10 | 4 | 7 | 0.57 |
| 15 | 2.5 | 4.5 | 0.56 |
| 20 | 2 | 3.5 | 0.57 |

It can be noted (Fig. 2, Table 2) that with an increase in the primary electrons energy, the scattering radius decreases both in the case of using titanium and in the case of using Kapton®. However, in the case of using the Kapton® film, we have a much smaller scattering radius. For example, the scattering radius for 10 MeV electrons when using Kapton® is 4 mm (Table 2). This value is even less than the scattering radius of 15 MeV electrons in the case of using the titanium film. The scattering radius of 15 MeV electrons is 4.5 mm.

It can be noted that the general tendency towards a decrease in the electrons scattering radius persists with an increase in the electron energy from 3 MeV to 20 MeV (Table 2). The scattering radius of electrons at the distance of 20 cm after passing through the Kapton® film is smaller than the scattering radius of electrons after passing through the titanium foil. Therefore, we can assume that the use of the Kapton® film instead of the titanium foil in the design of the linear electron accelerators output devices would make it possible to obtain narrower electron beam.

One of the tasks of this work is to determine the possibility of reducing the bremsstrahlung background after passing through the output device of the electron accelerator. The energy spectra of bremsstrahlung gamma quanta at the distance of 20 cm after passing through the titanium foil, as well as after the Kapton® film, are shown in Figure 4. The calculation step for each spectrum is 100 keV.

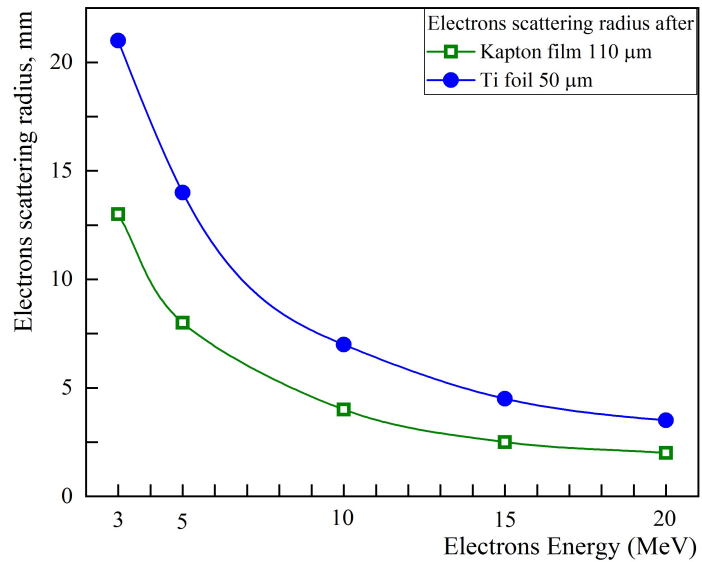

**Figure 3.** The comparison of the electrons scattering radii at the distance of 20 cm after the titanium foil (blue solid dots) as well as after the Kapton® film (green open points)

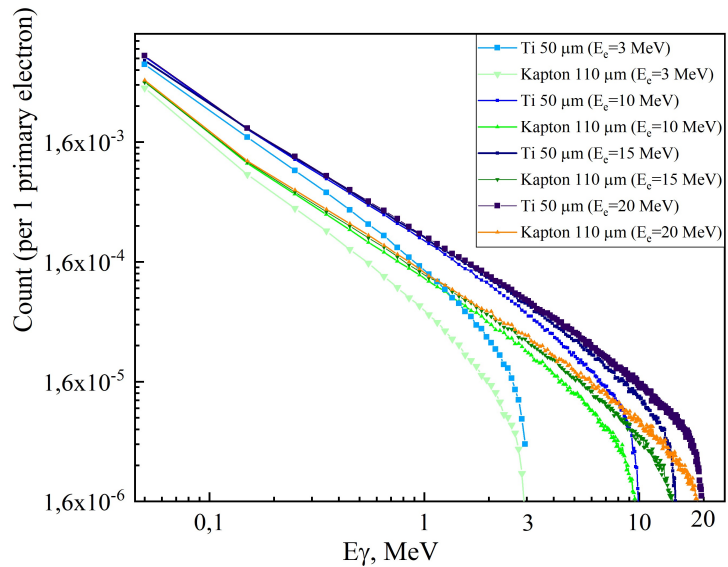

**Figure 4.** The comparison of the bremsstrahlung spectra at the distance of 20 cm after the titanium foil as well as after the Kapton® film for different values of the primary electrons energies

It can be seen (Figure 4) that the amount of bremsstrahlung gamma quanta at the distance of 20 cm after a 110 μm thick Kapton® film is less than the amount of bremsstrahlung gamma quanta after a 50 μm thick titanium foil for the same values of the primary electrons energy. The numerical values of the bremsstrahlung gamma quanta amount for each material, as well as the coefficient of their ratio, are presented in Table 3.

**Table 3.**
The numerical values of the bremsstrahlung gamma quanta amount (S) for each material, as well as the coefficient of their ratio, at the distance of 20 cm after Kapton® or titanium

| Electrons energy, MeV | S (for Kapton®) | S (for Ti) | S(Kapton®) / S (Ti) |
|---|---|---|---|
| 3 | 452486 | 804833 | 0.562 |
| 10 | 634884 | 1.148×10$^6$ | 0.553 |
| 15 | 676580 | 1.198×10$^6$ | 0.564 |
| 20 | 724679 | 1.309×10$^6$ | 0.554 |

Figure 5 shows the values of the bremsstrahlung amount at the distance of 20 cm after the foil. Data are normalized to 1 primary electron.

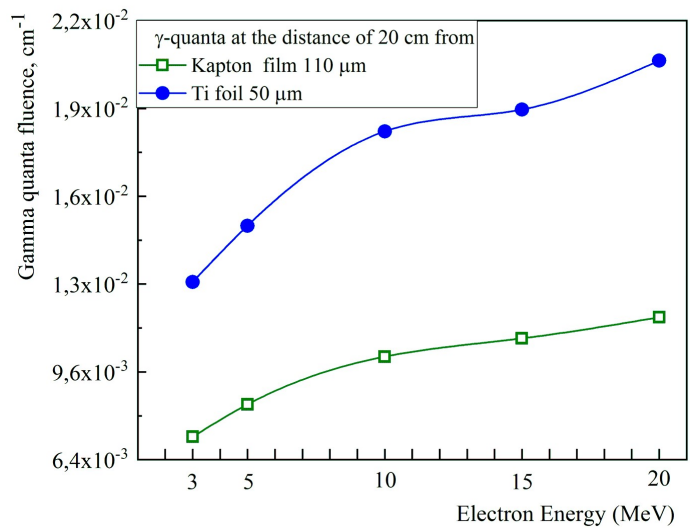

**Figure 5.** The comparison of the bremsstrahlung fluences at the distance of 20 cm after the titanium foil (blue solid points) as well as after the Kapton® film (green open points)

Gamma quanta are formed as a result of the primary electrons passage through the titanium foil, or the Kapton® film (Figure 5). It can be noted that in the case of using Kapton®, the fluence of bremsstrahlung gamma quanta is significantly less.

## CONCLUSIONS

A series of simulations of the electron beam passage through 50 μm thick titanium foil, as well as through 110 μm thick Kapton® film, are carried out for electrons of various energies. The electron energies are 3 MeV, 5 MeV, 10 MeV, 15 MeV, and 20 MeV. The fluence values of bremsstrahlung gamma quanta after the passage of the foil as well as after the film are obtained as a result of simulation. The most probable values of the absorbed energy in the titanium foil and in the Kapton® film are calculated for each value of the primary electrons energy. The values of the most probable scattering radius of an electron beam after passing through a 50 μm thick titanium foil and after passing through a 110 μm thick Kapton® film were calculated for the same values of the primary electrons energy. A comparative analysis of the obtained results of computational experiments is carried out. It is shown that the ratio of the total amount of bremsstrahlung gamma quanta when using a Kapton® film is approximately 0.56 to the total amount of bremsstrahlung gamma quanta when using a titanium film. The ratio coefficients of the most probable values of the scattering radii of electrons at a distance of 20 cm after passing through Kapton® to those after titanium are from 0.62 at 3 MeV electrons energy to 0.57 at 20 MeV electrons energy.

Analysis of the calculated data showed that the use of Kapton® ($C_{22}H_{10}N_2O_5$) as a material for the manufacture of high-energy electron beam output devices is promising in comparison to titanium films. The use of Kapton® instead of titanium makes it possible to significantly reduce the background of the produced bremsstrahlung gamma rays, as well as to reduce the scattering radius of the electron beam. Additional studies of the material thermal stability are necessary for the final decision on the experimental research.

**ORCID IDs**
**Tetiana V. Malykhina**, https://orcid.org/0000-0003-0035-2367; **Stepan G. Karpus**, https://orcid.org/0000-0002-1087-9245
**Oleg O. Shopen**, https://orcid.org/0000-0003-3158-8081; **Valerii I. Prystupa**, https://orcid.org/0000-0002-0789-2991